\documentclass[printer]{aa}
\usepackage{epsfig}
\usepackage{graphicx}
\usepackage{amssymb}
\usepackage{natbib}
\def\xte{XTE\,J1810$-$197}

\def\e15{1E\,1547.0-5408}

\def\sgrn{\mbox{SGR~0501$+$4516}}
\def\sg16{SGR\,1627-41}

\def\cxou{CXOU J164710.2$-$455216}

\def\swift{{\em Swift}}

\def\XMM{{\em XMM-Newton}}
\def\CXO{{\em Chandra}}

\def\ss{s\,s$^{-1}$}
\def\ergscm{\rm erg\,cm^{-2}\,s^{-1}}
\def\INT{{\em INTEGRAL}\,}

\begin{document}

\title{Multi-instrument X-ray monitoring of the January 2009 outburst from the recurrent magnetar candidate \e15}

\author{F. Bernardini\inst{1,2}
%{\email{bernardini@oa-roma.inaf.it}}
\and G. L. Israel\inst{2}
%{\email{gianluca@oa-roma.inaf.it}}
\and L. Stella\inst{2}
\and R. Turolla\inst{3,4}
\and P. Esposito\inst{5}
\and N. Rea\inst{6}
\and S. Zane\inst{4}
\and A. Tiengo\inst{7}
\and S. Campana\inst{8}
\and D. G\"otz\inst{9}
\and S. Mereghetti\inst{7}
\and P. Romano\inst{10}
%{\email{stella@oa-roma.inaf.it}}}
}

\offprints{F. Bernardini: bernardini@oa-roma.inaf.it}
\titlerunning{Jan 2009 outburst from magnetar candidate \e15}
\authorrunning{Bernardini et al.}

\institute{Universit\`a degli Studi di Roma ``Tor Vergata"
Via Orazio Raimondo 18, I$-$00173 Roma, Italy
\and INAF $-$ Osservatorio Astronomico di Roma, Via Frascati 33,
I$-$00040 Monteporzio Catone (Roma), Italy.
\and Department of Physics, University of Padova, Via Marzolo 8, I$-$35131 Padova, Italy
\and Mullard Space Science Laboratory
University College London,
Holmbury St Mary, Dorking, Surrey, RH5 6NT, UK
\and INAF - Osservatorio Astronomico di Cagliari, Localit\'a Poggio dei Pini, strada 54, I-09012, Capoterra, Italy
\and Institut de Ciencies de l'Espai (ICE, CSIC-IEEC), 08193, Barcelona, Spain
\and INAF - Istituto di Astrofisica Spaziale e Fisica Cosmica Milano, Via Edoardo Bassini 15, 20133 Milano, Italy
%\and Dipartmento di Fisica ``Enrico Fermi'', Universit\'a di Pisa,
%Largo B. Pontecorvo 3, I$-$56127 Pisa, Italy
\and INAF $-$ Osservatorio Astronomico di Brera, Via Bianchi
    46, I$-$23807 Merate (Lc), Italy
\and AIM (UMR 7158 CEA/DSM-CNRS-Universit\'e Paris Diderot) Irfu/Service d’Astrophysique, Saclay, F-91191 Gif-sur-Yvette Cedex
\and INAF - IASF Palermo Via Ugo La Malfa 153, 90146 Palermo, Italy}
\date{}

\abstract
   {With two consecutive  outbursts recorded  
in four months (October 2008 and January 2009), and a possible third outburst in 2007,
\e15 is one of the most active transient anomalous X-ray pulsars known so far.}
   {Thanks to extensive X-ray observations, obtained both in the quiescent and active states, \e15 
represents a very promising laboratory to get insights into
the outburst properties and magnetar emission mechanisms.}
   {We performed a detailed timing and spectral analysis of four \CXO, three \INT, 
 and one \XMM\ observations collected over a two week interval after the outburst onset 
in January 2009. Several \swift\ pointings, covering a 1.5 year interval, 
were also analyzed in order to monitor the decay of the X-ray flux.}
   {We compare the characteristics of the two outbursts, as well as those of the active
and quiescent states. We also discuss the long-term X-ray flux history
of \e15\ since its first detection in 1980, and show that the source 
displays three flux levels: low, intermediate and high.}
   {}


\maketitle

\section{Introduction}
Anomalous X-ray Pulsars (AXPs) and Soft Gamma-ray Repeaters
(SGRs) are young ($\sim10^{4}$~yr), isolated neutron stars (NSs)
whose X-ray luminosity greatly exceeds their rotational energy
losses. Both classes of objects show pulsations in the X-ray
band, with spin period clustering in the 2$-$12~s range and period
derivatives $\dot{P}\sim10^{-10}$--$10~^{-13}$\ss. The dipole magnetic field strength, 
as inferred via the standard formula, is $B\sim10^{14}$--$10^{15}$ G. 
There is a wide consensus that the activity of these sources is sustained by the
rearrangement/decay of the extremely strong magnetic field in their interior (the
magnetar model; Duncan \& Thompson 1992; Thompson \& Duncan 1995).

To date, there are 18 confirmed magnetars (11 AXPs and 7 SGRs) plus
a few additional candidates (for a review  see Mereghetti 2008)\footnote{see
also {\tt http://www.physics.mcgill.ca/\textasciitilde
pulsar...\\.../magnetar/main.html} for an updated catalog of SGRs/AXPs.}.
Ordinarily divided in two classes, 
there is now increasing evidence that the distinction between 
AXPs and SGRs originates mainly from the way in
which the sources are first discovered (rather than reflect intrinsic
physical differences, as also supported by
recent MHD simulations, Perna \& Pons 2011): AXPs, are detected by their persistent pulsed
emission in the X-ray band, and SGRs are discovered through the emission of
short, repeated bursts of hard X-ray/soft gamma-rays. However, SGR-like bursts
have now been detected from several AXPs, and persistent pulsed X-ray
emission has been observed from all SGRs.  

AXPs and SGRs display X-ray variability which extends over 
several orders of magnitude in both intensity and timescale:
from slow and moderate flux changes (up to a factor of a few) on
timescales of years (shown by all members of the class), to
moderate/intense outbursts (flux variations of a factor up to 10)
lasting 1--3 years (1E\,2259$+$586, and 1E\,1048.1$-$5973), to
dramatic and intense SGR-like burst activity on sub-second
timescales (4U\,0142$+$614, XTE\,J1810$-$197, 1E\,2259$+$586, and
1E\,1048.1$-$5973, besides all the SGRs; see e.g. Kaspi et al. 2007). 
Furthermore, in 2003 the first Transient Anomalous X-ray Pulsar (TAXP), \xte, was discovered
(Ibrahim et al. 2004). The source was   
one of thousands of faint ROSAT X-ray sources; it
suddenly displayed a strong flux increase (factor of about 100), 
which allowed the detection and measurement to measure of 
P and $\dot{P}$ and revealed its magnetar nature. Thanks to the high flux level, it was
possible to follow evolution of the the timing and spectral properties for
several years after the outburst: this has provided the most extensive coverage 
of a transient magnetar from outburst to quiescence so far
(Bernardini et al. 2009). In the last few
years, six other faint X-ray sources underwent similar outbursts
(X-ray flux variation of a factor $\sim100$). 
These sources were consequently classified as transient magnetars: 
\e15, \cxou, \sg16, \sgrn, \e15, SGR\,0418$+$5729, and SGR\,1833$-$0832 
(Muno et al 2007; Israel et al. 2007, Esposito et al. 2008; Rea et al. 2009; 
van der Horst et al. 2010, G{\"o}{\u g}{\"u}{\c s} et al. 2010, Esposito et al. 2010a,b). 
This suggested that presently known sources constitute only a
fraction of a much larger, still undetected, magnetar population.

Here we present a multi-instrument X-ray monitoring of the January 2009 
outburst of the transient magnetar \e15. Results are compared with those of the October 2008 
outburst, as well as results from archival data since the first source detection in 1980. 

\section{1E 1547.0-5408: discovery and previous X-ray campaigns}

\e15 (known also as SGR 1550-5418, see i.e. Rea et al. 2008) was discovered in 1980 with the Einstein X-ray
satellite (Lamb \& Markert 1981), and then studied in detail for
the first time by Gelfand \& Gaensler (2007) with an \XMM\ observation 
carried out in 2006. These authors proposed the source as a magnetar 
candidate, based on its spectrum composed by the sum of a blackbody 
(BB) plus a powerlaw (PL), like many other magnetar candidates, 
and on a possible association with the young
supernova remnant G327.24-0.13.

On June 22, 2007 the \swift\ satellite caught the source at an
X-ray flux a factor $\sim16$ times higher than that previously
recorded by \XMM\ in August 2006: $F^{June \, 07}_{1-8\,keV} \sim
5 \times10^{-12}\,\ergscm$, as compared to $F^{Aug \, 06}_{1-8\,keV}
\sim 3 \times10^{-13}\,\ergscm$ (Gelfand \& Gaensler 2007,
Halpern et al. 2008). No magnetar-like bursts were observed, possibly due
to the sparse X-ray coverage.

\e15 is one of two sources in the magnetar class (the other is \xte), which 
showed transient pulsed radio emission during its outburst 
(Helfand et al. 2006, Camilo et al. 2006, Camilo et al. 2007, Burgay et al. 2009). 
Using data collected in June 2007 
with the Parkes radio telescope and the Australia
Telescope Compact Array, Camilo et al. (2007) unambiguously
revealed the magnetar nature of the source, by measuring the spin
period and period derivative, $P\sim2.069$~s and
$\dot{P}\sim2.3\times10^{-11}$~\ss. \e15 was undetected in
previous archival radio observations (starting from 1998),
implying a flux at least 5 times lower then that recorded in 2008, 
and consequently suggesting a transient behaviour for the source also at radio
wavelengths ($F_{1.4\rm\,GHz}^{2008}=2.5\pm0.5$ mJy, 
$F_{1.4\rm\,GHz}^{1998}\leq0.5$ mJy). The source distance derived from the dispersion 
measure (Camilo et al. 2007) was $\sim9$~kpc, larger 
than the value of 4--5~kpc previously proposed by
Gelfand \& Gaensler (2007) on the basis of a possible
association with G327.24$-$0.13. 

Following the relatively deep  \XMM\ pointing taken in 2006
during quiescence, a second observation was carried out in 2007,
during outburst decay. Both spectra were successfully fit 
with a BB plus PL model. In the former 
$kT_{BB} \sim 0.40$~keV, $\Gamma \sim 3.2$ (Halpern
et al.~2008), while the latter observation was characterized by a harder
emission, with $kT_{BB} \sim 0.52$~keV and $\Gamma \sim 1.8$. 
Here $kT_{BB}$ is the BB temperature and $\Gamma$ the photon index of the PL. 

The \XMM\ X-ray data taken in 2007 were found to be weakly
modulated, with a pulse fraction (PF) of about $7\%$, one of
the lowest ever recorded in magnetar candidates.  The pulse shape was
complex, showing indications of variability both with energy
and flux. Only a marginal detection of pulsations was
reported in the \XMM\ observation of August 2006: the X-ray PF was
$\sim15\%$ (Halpern et al. 2008), a value consistent with the
upper limit previously derived by Gelfand \& Gaensler (2007) on
the same data set.

\subsection{Confirmed outbursts}

\begin{figure*}[!]
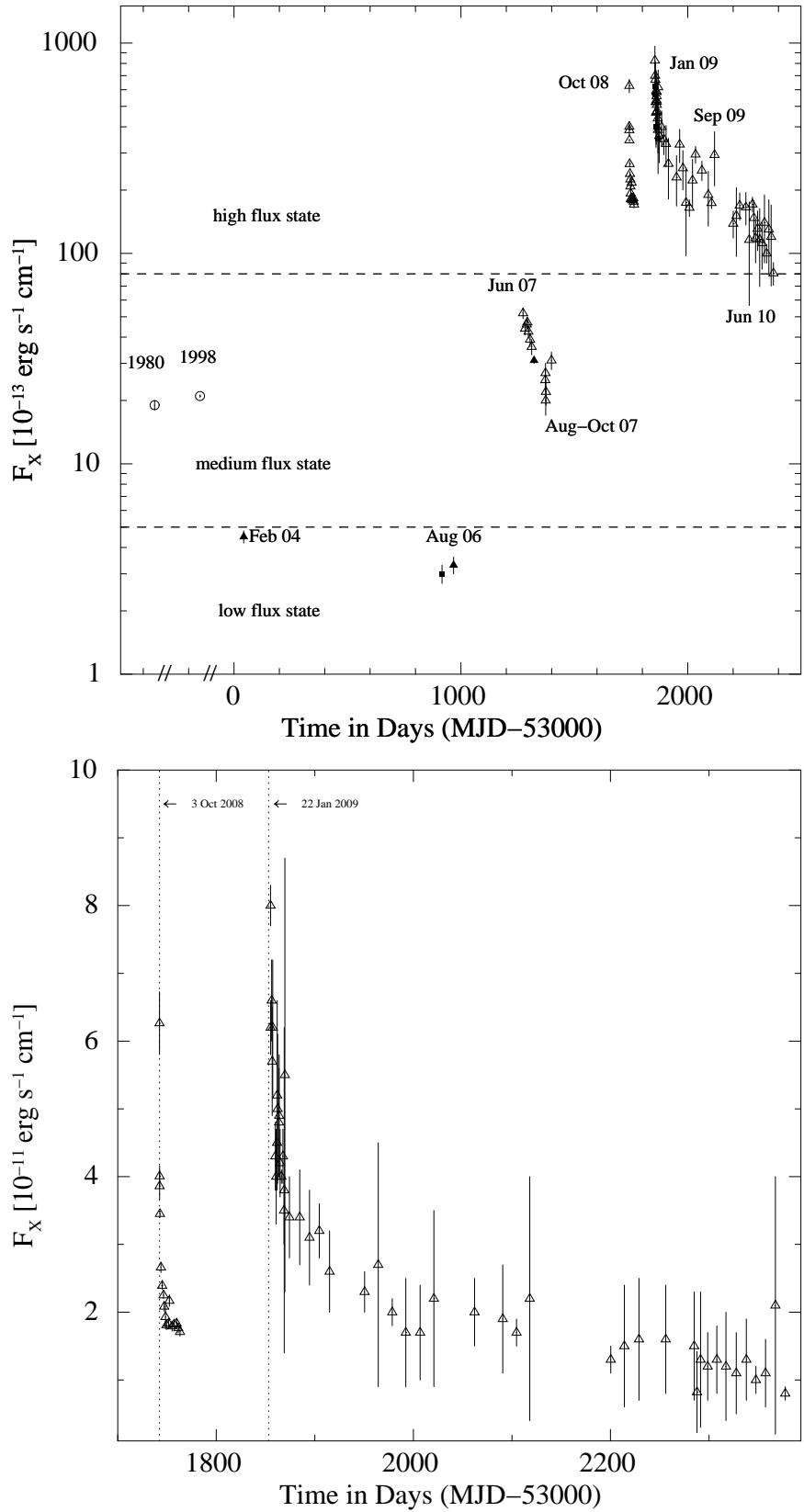

\begin{center}
\includegraphics[width=0.58\textwidth, angle=-90]{fluxhistfinal.ps}
\includegraphics[width=0.60\textwidth, angle=-90]{fluxev_2outb.ps}
\caption{\textit{Upper panel}: X-ray flux vs time. Empty triangles are 2$-$10 keV \swift\ data,
black triangles are 0.5$-$10 keV \XMM\ data, while black squares are 0.5$-$10 keV \CXO\ data, 
blank circles are 0.5$-$10 keV Einstein 1980 and $ASCA$ 
1998 data (the latter two values are from Gelfand \& Gaensler 2007). 
The X axis below the zero value displays two discontinuity in order to easily 
compare the recorded fluxes with that of Einstein (1980) and $ASCA$ (1998). The empirically selected 
horizontal dashed lines highlight the distinction among different flux states (see section \ref{section:spec} and 
\ref{3states} for details). 
%which was $1.9\pm^{1}_{0.5}\times10^{-12}\ergscm$ 
%and $2.1\pm0.3\times10^{-12}\ergscm$ respectively; Gelfand \& Gaensler, 1998; Halpern et al., 2008). 
\textit{Lower panel}: 2$-$10 keV flux for the October 2008 (Israel et al. 2010) 
and January 2009 outburst. Dotted vertical lines represent the two outbursts 
trigger time. Errors in both panels are $1\sigma$ c.l.. All reported fluxes are not corrected for absorption.}
\label{fig:fluxhistory}
\end{center}
\end{figure*}

\e15 represents a rare case among magnetars: it showed two
consecutive outbursts (with X-ray flux variation $>160$) within 
a few months (October 2008, and January 2009), 
and likely a third one (for
which the beginning phase was missed) occurred sometime before
June 2007, just one year before the first confirmed outburst 
(see Figure \ref{fig:fluxhistory} for a summary of the
available X-ray observations of \e15 since January 1980).

\subsubsection{The October 2008 outburst}

On October 3, 2008, \e15 entered an outburst state, exhibiting a
series of short bursts accompanied by a strong increase 
in the persistent X-ray flux. Thanks to the prompt response of the
\swift\ observatory, the source was monitored starting from only
$\sim100$ s after detection of the first burst. The maximum
flux in the 2$-$10 keV band was found to be $6.3\pm0.5\times10^{-11}\,\ergscm$ (Israel
et al. 2010), i.e. $\sim160$ times higher than its historical minimum
level of August 2006 (see Figure \ref{fig:fluxhistory}). 
During the three weeks of \swift\ monitoring after the outburst 
onset (total of 17 pointings), the X-ray flux
was found to decay following a powerlaw of index
$\alpha\sim-0.17$, reaching a flux of 
about $\sim1.5\times10^{-11}\,\ergscm$ (three weeks after the outburst onset). 
Israel et al. (2010) found that the outburst spectrum 
could be modeled with a thermal (BB)
plus a non-thermal (PL) component as 
often the case in magnetar candidates (Mereghetti et al. 2008). 
In particular, the spectrum was initially dominated by an hard 
PL with $\Gamma \sim1.1$; later, while the flux decreased, it became softer and a BB
component ($kT \sim 0.75$~keV) became dominant. Moreover, Israel et al. (2010) 
found that the PF increased, from 20\% to 50\% on a 21 days baseline, 
following the outburst onset in October 2008. Over that baseline 
these authors found a phase coherent timing solution with 
$\dot{P}=3\times10^{-11}$ s/s, and $\ddot{P}=2\times10^{-17}$ s/s$^2$. 

\subsubsection{The January 2009 outburst}

On January 22, 2009 (MJD=54853.037) the source entered a new state of bursting
activity (discovered by \swift\ and $Fermi$; Gronwall et al. 2009, Connaughton 
\& Briggs, 2009), characterized by a strong X-ray
flux increase, which culminated when more than 200 bursts were
recorded by the $INTEGRAL$ satellite in a few hours (Mereghetti et
al. 2009). A new X-ray monitoring campaign was initiated, involving
a number of high-energy observatories, including \XMM, \CXO, $INTEGRAL$,
\swift, Suzaku, and $Fermi$. Among other things, this led to the
spectacular discovery of multiple expanding rings surrounding the image 
of the X-ray source. These rings were caused by
scattering of the photons emitted by the AXP during  a bright 
burst on January 22, 2010 off different layers of interstellar 
dust (Tiengo et al. 2010); this yielded an estimate 
of the source distance, which turned out to be $\sim 4$--5~kpc. 

\section{Observations and data analysis}
\label{obs}
Here we report on four \CXO\,, one \XMM\,, and three $INTEGRAL$
pointings of \e15, carried out after the outburst onset of January
22, 2009 and covering a total baseline of 15 days. In order to
follow the X-ray flux evolution over a longer period, we also
analyzed 44 \swift\ observations, covering about 1.5 years after the
outburst. Data from Suzaku and $RXTE$ were also used 
in order to get a phase coherent 
timing solution over a 15 day baseline. 

We compare the results of our timing and spectral
analysis with those available in the literature in relation to its previous 
outbursts and states of low activity.
We then study the evolving spectrum within the framework 
of the twisted-magnetosphere model.
The analysis of the burst emission detected by $INTEGRAL$ has been presented separately
by Mereghetti et al. (2009), see also Savchenko et al. (2010). 
\subsection{Chandra and XMM-Newton}

\CXO\ observed the source four times, all in Continuous Clocking
(CC) faint mode. The first pointing was carried out on Jan
23, 2009 ($\sim2$ days after the outburst onset), and lasted 10 ks; it
was the only one made using the HETG in front of the ACIS-S CCD.
The second observation was carried out on Jan 25, 2009 (12~ks), the third
on Jan 29, 2009 (13~ks), and the last one on  Feb 06, 2009 (15
~ks). The total monitoring interval was about 15 days.
\XMM\ observed the source for $\sim$58~ks on Feb 03, 2009, with both the
\textit{pn} and MOS1/2 cameras in Full Frame mode and with the thick filter
applied. \CXO\ and \XMM\ data were reprocessed using CIAO 4.2 and SAS 
(9.0.0), respectively; in both cases we used the latest version of the
calibration files available at the time of the analysis.

\CXO\ CC-mode lightcurves and source/background spectra were extracted 
using  $dmextract$ from regions $50\arcsec$ wide.
The background region was selected far enough from the source in order 
to exclude contamination by the three expanding X-ray scattering rings 
(for the ring position with time see Tiengo et al. 2010). 

Both \XMM\ spectra and lightcurves were extracted using
a circular region of radius 55\arcsec, enclosing $\sim90\%$ of the source
photons, (no significant pile up was detected). 
A background region of the same size was selected in
the same CCD in which the source lied, in order to avoid the three
X-ray scattering rings. Source events were selected and filtered so as
to remove any possible rapid (t$<1$ s) burst contamination. 
All spectra were analyzed using the latest version of $XSPEC$ (12.5.1n).

\subsection{Swift}
44 \swift\ observations, in both photon counting (PC) and
windowed timing (WT) readout modes, were analyzed. 
In PC mode the entire CCD is read every 2.507 s, while in WT mode
only the central 200 columns are read while one-dimensional imaging is
preserved, achieving a time resolution of 1.766 ms. The
\swift\ observations
span the period from Jan 23, 2009, to June 
30, 2010, totaling a  net exposure of $\sim90.0$ ks and $\sim38$ ks
in WT and PC modes, respectively.

The raw data were processed with \textsc{xrtpipeline} (version 0.12.3, in
the \textsc{heasoft} software package 6.6), standard filtering and screening criteria were applied by using
\textsc{ftools} tasks. We accumulated the PC source events from a circle
of 20 pixels radius ($\sim90\%$ of source photons; one pixel corresponds to about 2.36\arcsec)
and the WT data from a $40\times40$ pixels box along the image strip. To
estimate the background, we extracted PC and WT events from source-free
regions far from the position of \e15.
For the spectral fitting, the ancillary response files (arf) were
generated with \textsc{xrtmkarf}; they account for different
extraction regions, vignetting and point-spread function corrections. We
used the latest available spectral redistribution matrix (rmf) in
\textsc{caldb} (v011).

In the context of the present work,
the spectral analysis of the \swift\ data is 
mainly aimed at obtaining long-term
flux measurements for \e15, after its January 2009 outburst.

\subsection{INTEGRAL}
The source was observed by \INT\ during orbits 767$-$771, from
Jan 24, 2009, to Feb 4, 2009. These data have been obtained
through of a public ToO programs. 
We analyzed the IBIS/ISGRI (Ubertini et al. 2003;
Lebrun et al. 2003) data by using the spectral-imaging technique of
G{\"o}tz et al. (2006). The source flux was determined in narrow energy
bands through mosaicked images of individual pointings (typically lasting
45 minutes), which were then used to build spectra to be
fitted with the correspondingly rebinned response matrix.  To build
our spectra, we chose 7 energy bands: 18$-$25, 25$-$40, 40$-$60,
60$-$80, 80$-$100, 100$-$150, and 150$-$300 keV.

\section{Results}
\label{results}

\subsection{Timing analysis}
\label{sec:timing}
\begin{figure}
\begin{center}
\includegraphics[scale=.5, angle=-90]{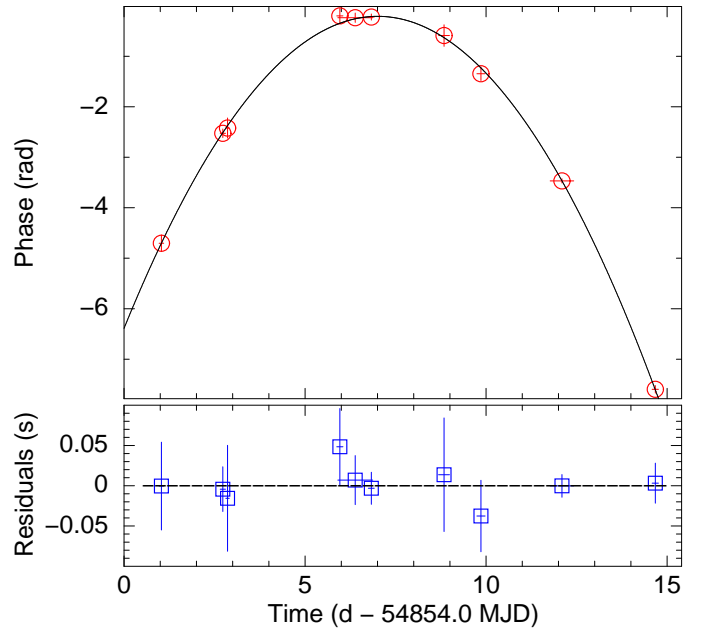}
\caption{\CXO, \XMM\, $Swift$, $RXTE$ and Suzaku pulse phase
evolution with time, together with the time residuals with respect to the unique phase-coherent
timing solution discussed in the text (shown by the solid line in the upper panel).}
\label{fig:phase}
\end{center}
\end{figure}

\begin{figure*}
\begin{center}
\includegraphics[scale=.6]{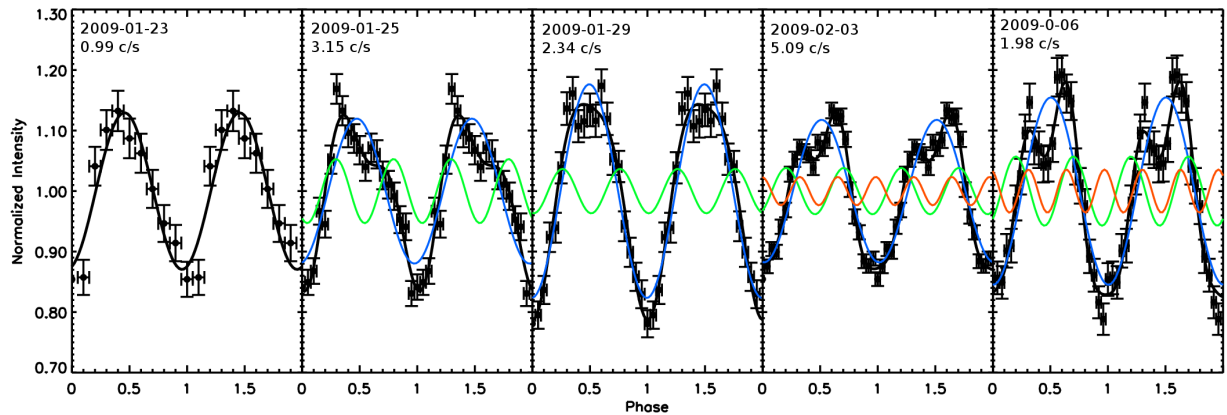}
\caption{0.5$-$10 keV pulse profile evolution
in time. Black: data and best fitting model. The different
harmonics contributing to the best fit model are also shown: blue, green
and red curves are the first, second and third harmonic, respectively (see
text for details). The background subtracted average count rate is also reported 
in each panel. The low count rate of the Jan 23, 2009 observation is due to the presence 
of the grating in front of the CCD, while the Feb 03, 2009 observation 
was performed with \XMM, which has a larger effective area with respect to \CXO.}
\label{fig:pulse}
\end{center}
\end{figure*}

In order to measure the timing properties of \e15\ and carry out a
phase-coherent pulse phase spectroscopic (PPS) study of the Chandra and
\XMM\ observations we performed a detailed timing analysis of all
the available archival X-ray datasets including \swift, $RXTE$ and Suzaku
observations. We used the 1--10 keV band for all instruments
but $RXTE$, for which we used the 2--10 keV band. 
Photon arrival times  were corrected to the
barycenter of the Solar system with the \textsc{barycorr} task 
(we used RA=15$^h$ 50$^m$ 54$\fs$12, Dec=-54$^{\rm o}$ 18$^{\prime}$
24$\farcs$19 and J2000 for the source position; Israel et
al. 2010) and by using the same ephemeris file (DE200) and coordinate
reference system (FK5) for all observations.
We first derived an accurate period measurement by folding the
data from the Suzaku pointing 
(which has the longest baseline, see Enoto et al. 2010)
and, subsequently, we studied the phase evolution across different
observations by means of a phase-fitting technique 
(details on our adopted technique are in Dall'Osso et al. 2003). 
Given the complex (double-peaked) and highly variable pulse shape, 
we fitted the lightcurve from each observation
with a Fourier sine series truncated at the latest significant harmonic 
(see Table \ref{tab:tabchi} for the fit results).
Indeed, the third harmonic became statistically significant in the 
fit only during the last two pointings. 
The second harmonic was always statistically
significant (more than $3\sigma$ c.l.), with the
exception of the first pointing in which only the first harmonic was
significant (possibly owing to the reduced signal to noise, S/N, ratio). 
The statistical significance of the inclusion of higher harmonics 
respect the fundamental one was evaluated by an F-test 
(see Table \ref{tab:tabchi}). 
A best fit ($\chi^2$=0.9 for 7 degree of freedom, d.o.f.) 
phase coherent timing solution (reported in Figure\,\ref{fig:phase}), 
could be determined unambiguously and contained only   
the P and $\dot{P}$ terms. This timing solution gave 
$P=2.0721257(3)$\,s and $\dot{P}=2.27(3)\times 10^{-11}$\,s\,s$^{-1}$
($\nu=0.48259620(6)$\,Hz, and $\dot{\nu}=-5.29(6)\times10^{-12}$ Hz\,s$^{-1}$; 
epoch 54854.0 MJD; valid between 54854.0 and 54869.0 MJD). 
Here and thereafter 1$\sigma$ c.l uncertainty is reported, unless 
otherwise specified. These values are consistent with those
reported by Kaneko et al. (2010) and Ng et al. (2010) based on Fermi
and $RXTE$ data only, respectively. Time residuals with respect to the 
timing solution are plotted in Figure\,\ref{fig:phase}. 
Their distribution clearly indicates that no higher derivatives 
of the period are required to fit the present data. By including 
$\ddot{P}$ component in the fit, we derived a $3\sigma$ 
upper limit of $1.8\times10^{-17}$ s/s$^2$ (absolute value). 
This is smaller then the $\ddot{P}$ component detected during 
the 2008 outburst (Israel et al. 2010; Ng. et al 2010). The timing 
solution obtained during 2008 outburst consequently resulted to be more complex then the 
2009 timing solution. 
The \CXO\ and \XMM\ resulting pulse profiles are shown in
figure \ref{fig:pulse}.  
The morphology of the pulse profile (0.5--10 keV band) evolved in time:
the first peak was clearly dominant in the second pointing, while the second peak
became dominant at later times. 

\begin{figure*}
\begin{center}
\includegraphics[angle=-90,scale=.50]{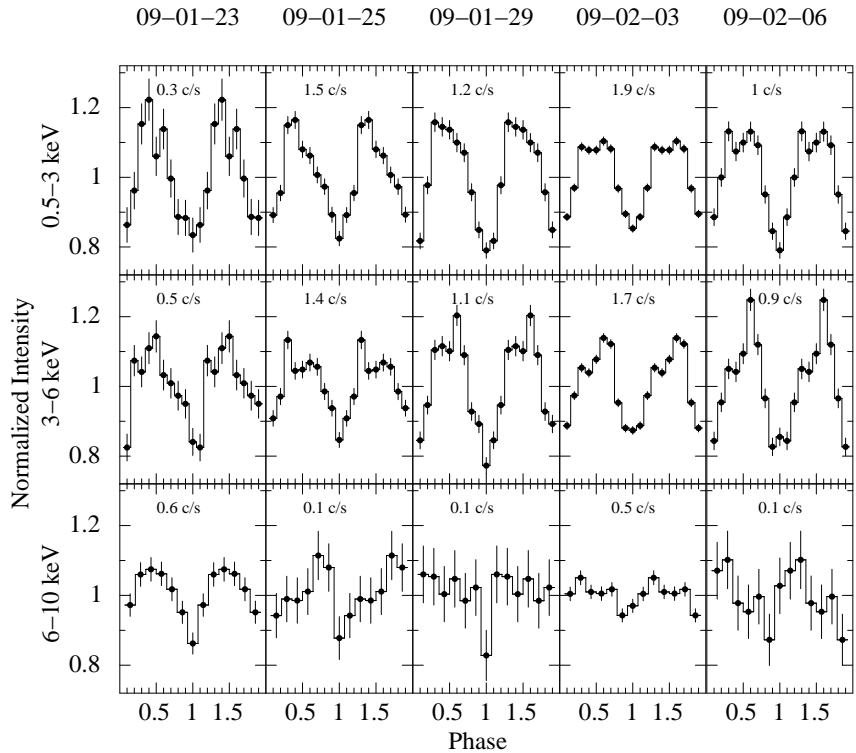}
\caption{0.5--3, 3--6, and 6--10 keV pulse profile evolution
with time (time increases from left to right). Energy increases 
from top to bottom. Background subtracted count rate is also reported.} 
\label{fig:pulprof}
\end{center}
\end{figure*}

In order to study the lightcurve evolution at different energies, we
divided the counts into three energy bands, 0.5--3, 3--6, 
and 6--10 keV (see Figure \ref{fig:pulprof}). 
Also the 0.5--3 keV and 3--6 keV pulse profiles 
were double-peaked and evolved from a configuration in which one peak 
was dominant to a configuration in which the other peak became dominant. 
The strength of the modulation is found to clearly decrease with energy.

\begin{table*}
\caption{Statistical significance ($\sigma$) for the inclusion of 
the second and the third harmonic during the five different pointings.
Data are in the 0.5$-$10 keV energy interval. $\chi^{2}$, and
degrees of freedom, are reported for each fit performed with 
$\rm I_{harm}$, $\rm I_{harm}+II_{harm}$, and $\rm I_{harm}+II_{harm}+III_{harm}$.}
\begin{center}
\begin{tabular}{cccccccccc}
\hline \hline
\\
Epoch & \multicolumn{3}{c}{$\rm I_{harm}$}&
\multicolumn{3}{c}{$\rm I_{harm}+II_{harm}$} & \multicolumn{3}{c}{$\rm I_{harm}+II_{harm}+III_{harm}$}\\
&$\chi^{2}$& d.o.f.&  $\sigma$    & $\chi^{2}$& d.o.f& $\sigma$ & $\chi^{2}$& d.o.f.& $\sigma$ \\
\hline \hline
Jan 23 2009 & 12 & 8 & $-$  & $-$ &  $-$& $-$ & $-$ & $-$& $-$
\\
Jan 25 2009 & 79 & 18 & $-$ & 25 & 16  & 3.8   &  $-$ & $-$& $-$
\\
Jan 29 2009 & 40& 18& $-$ & 18& 16     & 3.1   &   $-$ & $-$& $-$
\\
Feb 03 2009$^a$ &  179 & 23 &$-$ & 69 & 21 & 4.1    &  21 & 19 & 4.4
\\
Feb 06 2009 & 78 & 23 &$-$  &  36 & 21 & 3.7  & 17 & 19 & 3.3
\\
\hline \hline
\end{tabular}
\label{tab:tabchi}
\end{center}
$^a$ \XMM\ pointing.
\end{table*}

\begin{table*}
\caption{Root mean square pulsed fraction (PF), see text for details, in four energy intervals
(0.5$-$3~keV,
3$-$6~keV,
6$-$10~keV, and 0.5$-$10~keV). The last three columns show the
PF as computed separately, for the I, II, and III harmonic, in the
0.5$-$10 keV band (see text for
details). Uncertainties are $1\sigma$ c.l.}
\begin{center}
\begin{tabular}{cccccccc}
\hline \hline
\\
Epoch & PF$_{0.5-3.0\, keV}$ &  PF$_{3-6\, keV}$ & PF$_{6-10\,keV}$ & PF$_{0.5-10\, keV}$
& PF$^{Iarm}_{0.5-10\, keV}$ &  PF$^{IIarm}_{0.5-10\,keV}$ &  PF$^{IIIarm}_{0.5-10\, keV}$\\
    &\% &\% &\% &\% &\% &\%&\% \\
\hline \hline
Jan 23 2009& $   13\pm2  $  &  $ 9\pm2 $    & $ 7\pm2$    & $9\pm1$   &
$12\pm1.3
$  &  $ ^{b}<9    $  &  $  - $
\\
Jan 25 2009& $   11\pm1   $    &  $9\pm1 $    & $6\pm4 $    &  $9\pm1$  &
$11.7\pm0.7
$  &  $  5.4\pm0.7  $  &  $ ^{b}<2.5  $
\\
Jan 29 2009& $   14\pm1   $    &  $ 14\pm1$     & $ 5\pm4 $    &$14\pm1$
&   $17.7\pm0.8
$  &  $  3.7\pm0.8  $  &  $ ^{b}<3.7  $
\\
Feb 03 2009$^a$& $ 9.7\pm0.4     $  &  $ 10.0\pm0.4 $ & $ 3\pm1 $&   $9.0\pm0.3$  &
$11.8\pm0.3$ &
$  3.8\pm0.3  $  &  $ 2.4\pm0.4  $
\\
Feb 06 2009& $ 13\pm1   $    &  $14\pm1 $    & $ <16^{b}$    & $12\pm1$   &
$15.5\pm0.6
$  &  $  5.7\pm0.6  $  &  $ 3.6\pm0.6  $
\\
\hline \hline
\end{tabular}
\label{tab:tabpf}
\end{center}
$^a$ \XMM\ pointing.
$^b$ Upper limits are at $3\sigma$ c.l.
\end{table*}

We estimated the root mean square pulsed fraction, hereafter PF, which is defined as:\\
$${\rm PF} = \left( \frac{1}{\rm N} ( \sum_{i=1}^{\rm N}
({\rm R}_{\rm i}-{\rm R}_{\rm ave})^2 -
\Delta {\rm R}_{\rm i}^2) \right)^\frac{1}{2}/{\rm R}_{\rm ave}$$
where N is the number of phase bins (N=10 for 0.3--3, 3--6 and 0.5--10 keV energy intervals, and N=5 for 6--10 keV energy interval), 
${\rm R}_{\rm i}$ is the rate in each phase bin, $\Delta {\rm R}_{\rm i}$ is the 
associated uncertainty in the rate, and ${\rm R}_{\rm ave}$ is 
the average rate of the pulse profile.
Results are reported in Table \ref{tab:tabpf}. 
The PF resulted to decrease with energy. The drop was 
not very pronounced when comparing the low and medium energy
bands (0.5--3 and 3--6 keV), but was more significant in
the high energy band (6--10 keV). Instead, the PF evolution 
with time, within each energy band, does not show any clear trend, except for the high energy interval where PF was 
found to decrease with time  
(for a possible explanation of the PF time changes see 
\S\ref{rcs} and Figure \ref{fig:rvspf}).

For completeness, we also report, in Table \ref{tab:tabpf}, the PF as
computed separately for each of the harmonics that we used to represent the signal.
In this case the PF is defined as: $PF=(A_{max}-A_{min})/(A_{max}+A_{min})$, 
where $A_{max}$ and $A_{min}$ are the maximum and minimum value of the sinusoid respectively. 
Due to the lower S/N of the energy-resolved light curves, this procedure gave
meaningful results only for the (total) 0.5$-$10~keV energy band. We also
report the $3\sigma$ upper limit for the PF of the first statistically
non-significant harmonic.

\subsection{Spectral analysis}
\label{section:spec}

With three different recorded outbursts \e15\ is likely 
one of the most active transient magnetars.
To achieve a better understanding of the nature of this peculiar source, 
we began the analysis by collecting all the archival 
observations since the very first pointing made in
March 1980 by the Einstein satellite.

At present, three different flux levels were seen in \e15:
a low state, during the \XMM\ and \CXO\ pointings around August 2006 
($F_{X}\sim4\times10^{-13} \ergscm$);
an intermediate state, during the Einstein
1980 and $ASCA$ 1998 pointings,
and also during the \swift\ pointings performed 
between June 2007 and October 2007 ($F_{X}\sim2\times10^{-12} \ergscm$); and 
a high state, as seen during the two outbursts 
of October 2008 and January 2009 ($F_{X}\sim8\times10^{-11} \ergscm$). 
The recorded X-ray flux history (all reported fluxes are not corrected 
for absorption; see Figure \ref{fig:fluxhistory}) suggests that 
the source is highly variable and does not display a simple transient 
behaviour, with a single quiescent flux level. 
The term transient appears to reflect more the way in which the source was
discovered than its overall behaviour. 

\subsubsection{Black body plus powerlaw model}
\label{sub:bbpl}

\begin{figure*}
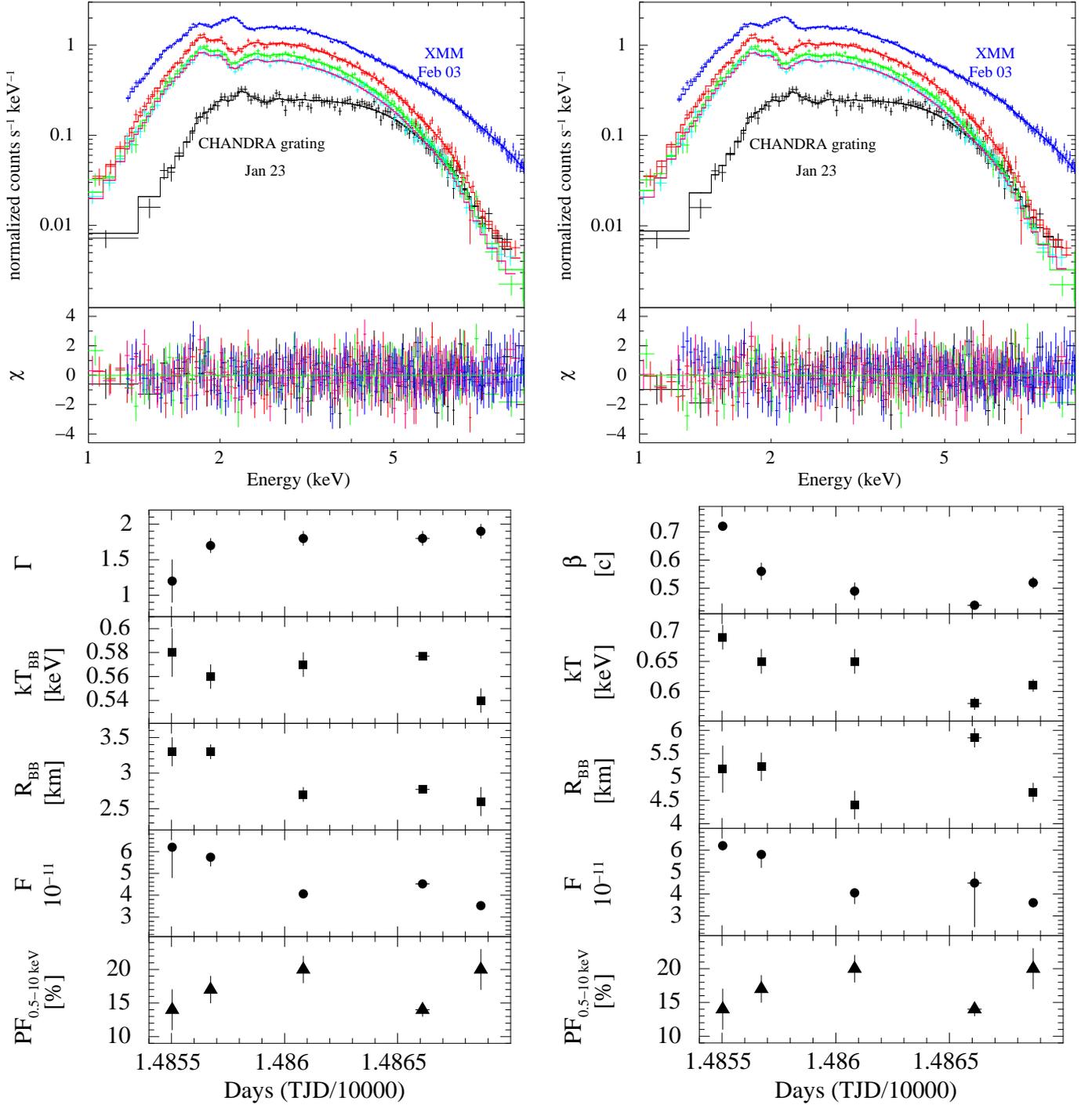

\begin{center}
\begin{tabular}{cc}
\includegraphics[width=0.45\textwidth, angle=-90]{bbplnormalres.ps} &\includegraphics[width=0.45\textwidth, angle=-90]{twistnormalres.ps}\\
\includegraphics[width=0.55\textwidth, angle=-90]{bbpl_left.ps} & \includegraphics[width=0.55\textwidth, angle=-90]{twist_right.ps}\\
\end{tabular}
\caption{\textit{Upper panels}: Count spectra and models in the 0.5--10 keV energy range for different epochs (black 
points and lines are Jan 23, 2009, red for Jan 25, 2009, green for 
Jan 29, 2009, blue for Feb 03, 2009, and magenta is Feb 06, 2009). Left: BB+PL model. 
Right: NTZ model. Fit residuals are shown in the bottom panels. 
\textit{Lower panels}: Time evolution of the best$-$fitting parameters inferred from the
BB+PL (left) and NTZ (right) fits of the 0.5$-$10~keV spectra. The 0.5--10~keV flux (in units of $10^{-11}\,\ergscm$)
and the PF evolution are also shown.}
\label{fig:speceeufs}
\end{center}
\end{figure*}

We began by applying the standard phenomenological AXP spectral model,
i.e. a BB plus a PL (a two component model is always required by the fit), to the 
0.5--10 keV spectrally resolved data from the January 2009 outburst.  
The fit was performed over the four \CXO\ and one \XMM\ data. 
All parameters were left free to vary with the only constraint that the
hydrogen column density remained the same at all epochs. All reported uncertainties 
hereafter are obtained by using the \textit{XSPEC} \textit{unc} command.
The results of this analysis are shown in Figure \ref{fig:speceeufs}, 
and reported in Table \ref{tab:spec}. Hereafter, the source distance is assumed to be 
4.5 kpc. We note that a significant excess in the \XMM\ PN fit residuals was detected below
1.2 keV, independent of the spectral model used. Similar residuals are
rather common in the PN spectra of bright and strongly absorbed sources,
suggesting that this soft excess is due to calibration issues (see, e.g., Boirin et
al. 2005; Sidoli et al. 2005, Martocchia et al. 2006). Consequently we analyzed the \XMM\
spectrum in the energy range 1.2--10 keV only. 

Based on these fits we found out that the outburst X-ray flux increase  
with respect to the recorded lower state of August 2006
($F^{Aug06}_{0.5-10\,keV}=3.3\pm^{0.1}_{0.3}\times10^{-13}\,\ergscm$
as compared to $F^{Jan09}_{0.5-10\,keV}=6.2\pm^{0.2}_{1.4}\times10^{-11}\,\ergscm$)  
was due to both a slight increase in the BB
temperature from $0.40\pm0.05$ keV to $0.58\pm0.02$ keV, 
and a hardening of the PL photon index, from $\Gamma=3.2\pm0.5$ to $\Gamma=1.2\pm0.3$.
Moreover, the spectral variation associated to the 
flux decay during the outburst (from
$F_{0.5-10\,keV}=6.2\pm^{0.2}_{1.4}\times10^{-11}\,\ergscm$ on Jan 23 to
$F_{0.5-10\,keV}=3.52\pm^{0.02}_{0.10}\times10^{-11}\,\ergscm$ on Feb 6) resulted from 
the decrease of both temperature and radius of the blackbody, 
and to the softening of the PL photon index. The BB temperature 
remained fairly constant during the first four pointings at an average 
$kT=0.57\pm0.01$ keV and afterwards it decreased slightly to
$kT=0.54\pm0.01$ keV, while the radius slightly decreased from $R_{bb}=3.3\pm0.2$ km to
$R_{bb}=2.6\pm0.2$ km. (the BB radius corresponds to distance of 4.5 kpc). 
The PL photon index also changes, becoming
softer, from $\Gamma=1.2\pm0.3$ to $\Gamma=1.9\pm0.1$ 
The $\chi^{2}_{red}$ of the joint fit is 0.97 (for 700 d.o.f.). 

\begin{table*}
\caption{Results from the simultaneous fit to all 0.5$-$10 keV spectra for the 
Jan-Feb 2009 observations. 1$\sigma$ c.l. uncertainties reported. 
BB+PL model: $N_H=3.46\pm0.03\times10^{22}\,{\rm cm}^{-2}$, $\chi^{2}_{red}=0.97$ for 700 (d.o.f). 
RCS model:   $N_H=3.06\pm0.02\times10^{22}\,{\rm cm}^{-2}$, $\chi^{2}_{red}=1.04$ for 700 (d.o.f). 
The source distance is assumed to be 4.5 kpc.}
\begin{center}
\begin{tabular}{cccccc}
\hline \hline
\multicolumn{6}{c}{BB+PL model}  \\
\hline
Epoch   &  $kT_{\rm BB}$   & $R_{\rm BB}$   & $\Gamma $      & \multicolumn{2}{c}{$F_{0.5-10\,keV}$}     \\
        &                  &                &                & \multicolumn{2}{c}{$\times10^{-11}$}         \\
        &   keV            &   km           &                & \multicolumn{2}{c}{erg\,cm$^{-2}$ s$^{-1}$}   \\
\hline
\\
Jan 23 2009&  $0.58\pm0.02 $  &  $3.3\pm0.2$   & $1.2 \pm0.3 $  & \multicolumn{2}{c}{$6.2\pm_{1.4}^{0.2}$}     \\
Jan 25 2009&  $0.56\pm0.01$   &  $3.3\pm0.1$   & $1.7 \pm0.1 $  & \multicolumn{2}{c}{$5.74\pm_{0.41}^{0.02}$}  \\
Jan 29 2009&  $0.57\pm0.01$   &  $2.7\pm0.1$   & $1.8\pm0.1$  & \multicolumn{2}{c}{$4.06\pm_{0.1}^{0.02}$}   \\
Feb 03 2009$^a$&  $0.580\pm0.003$ &  $2.77\pm0.05$ & $1.8\pm0.1$  & \multicolumn{2}{c}{$4.52\pm_{0.04}^{0.01}$}  \\
Feb 06 2009&  $0.54\pm0.01$   &  $2.6\pm0.2$   & $1.9\pm0.1$  & \multicolumn{2}{c}{$3.52\pm_{0.1}^{0.02}$}   \\
\\
\hline
\multicolumn{6}{c}{RCS model}    \\
\hline
Epoch    & $kT$               &$R$                &$\beta_{bulk}$      & $\phi$           & $F_{0.5-10\,keV}$     \\
         &                    &                   &                    &                  & $\times10^{-11}$         \\
         &   keV              & km                &                    & rad              & erg cm$^{-2}$ s$^{-1}$   \\
\hline
\\
Jan 23 2009&  $0.69 \pm0.02 $  & $5.2 \pm0.5 $     & $ 0.72 \pm0.01  $  & $0.47 \pm 0.01$    &$6.2  \pm0.2           $  \\
Jan 25 2009&  $0.65 \pm0.02 $  & $5.2 \pm0.3 $     & $ 0.56 \pm0.03  $  & $0.47 \pm 0.01 $   &$5.8  \pm_{0.6}^{0.1} $  \\
Jan 29 2009&  $0.65 \pm0.02 $  & $4.4 \pm0.3 $     & $ 0.49 \pm0.03  $  & $0.48 \pm 0.02 $   &$4.1  \pm_{0.5}^{0.02}$  \\
Feb 03 2009$^a$&  $0.58 \pm0.01 $  & $5.8 \pm0.2 $     & $ 0.44 \pm0.01  $  & $0.490\pm 0.005$   &$4.5  \pm_{2.0}^{0.5} $  \\
Feb 06 2009&  $0.61 \pm0.01 $  & $4.7 \pm0.2 $     & $ 0.52 \pm0.02  $  & $0.47 \pm 0.01 $   &$3.6  \pm0.2$  \\
\\
\hline \hline
\end{tabular}
\label{tab:spec}
\end{center}
$^a$ \XMM\ pointing.
\end{table*}

The evolution of the spectral parameters in the 2008 and 2009 outburst of \e15 
is difficult to compare. Indeed, only for the first three
\swift\ observations of the 2008 outburst a two component model
(BB+PL) is required (this might well be due to the lower S/N of the 
subsequent \swift\ observations), whereas a two component model is always required for the \CXO/\XMM\
data of the 2009 outburst. 
The 2008 analysis suggests that the PL is dominant in the first
pointing after the outburst onset and it is still detectable until the third pointing performed
one day after the outburst onset. This finding is not in contrast with the results of the 
\CXO/\XMM\ analysis of the 2009 outburst
which suggests a decrease in the PL photon index from the first
pointing of 23 Jan, 2009 ($\Gamma=1.2\pm0.3$) to the last one 6 Feb, 2009 ($\Gamma=1.9\pm0.1$).   

\subsection{Flux decay since Jan 23, 2009}
\label{subs:fluxdecay}

We adopted for all \CXO, \XMM, and \swift\ data
the same spectral decomposition: a BB plus a PL model,  
with the interstellar absorption, and fitted it to all 
the spectra together. All parameters were left free to vary, 
except for the absorption column density which 
was forced to be the same for among all datasets. 
Spectral fit were performed in the 2-10 keV range. 
This resulted in an acceptable fit ($\chi^2_\nu=1.07$ for 2601 d.o.f.). 
The fluxes derived in this way are plotted in Figure \ref{fig:fluxhistory}.
The 2$-$10 keV flux decreased from a maximum of
$8\pm1.4\times10^{-11}\,\ergscm$ to a minimum of
$8\pm1\times10^{-12}\,\ergscm$. 
The best fit model for the flux decay 
is a PL, $\propto (t-t_{0})^{-\alpha}$, 
with $\alpha=0.34\pm0.01$ ($\chi^2=0.92$ with 47 d.o.f). 

\subsection{Long term changes of intensity levels}
\label{3states}

\begin{table*}
\caption{Spectral parameters (BB+PL) from the three recorded 
source intensity levels. $1\sigma$ c.l. are reported.}
\begin{center}
\begin{tabular}{ccccccc}
\hline \hline
\\
state & $kT$  & $R$   & $\Gamma$  & $F_{0.5-10\,keV}$ & $L_{4.5}$ $^e$ & PF \\
      &  keV   &   km &           &  erg cm$^{-2}$ s$^{-1}$ & erg s$^{-1}$ &  \%         \\
\hline \hline
\\
\textbf{Low}$^{a}$ & $0.43\pm0.3 $  &  $0.7\pm0.2$  & $4.0\pm0.2$   & $3.7\pm^{0.1}_{0.3}\times10^{-13}$ & $9\pm^{0.2}_{0.8}\times10^{32}$ &$<15$ \\
\hline
\textbf{Intermediate}$^{b}$   & $0.52\pm0.01 $           &  $1.5\pm0.1$       & $3.0\pm0.4$   & $3.0\pm0.3\times10^{-12}$ & $7.3\pm0.5\times10^{33}$ &$\sim7$ \\
\hline
\textbf{High} \\
Minimum$^{c}$ & $0.69\pm0.02$            & $1.6\pm0.1$         & $5\pm1$      & $1.4\pm0.1\times10^{-11}$ & $3.4\pm0.2\times10^{34}$&$33\pm5$ \\
Maximum$^{d}$ &  $ 0.57 \pm 0.01   $  &  $  3.3\pm 0.1 $  & $1.48 \pm 0.03  $  & $5.8\pm0.7\times10^{-11}$  & $1.4\pm0.2\times10^{35}$& $10\pm1$\\
\hline \hline
\end{tabular}
\label{tab:activitystate}
\end{center}
$^{a}$ Values refer to the Aug 21, 2006 \XMM\ pointing.\\
$^{b}$ Values refer to the Aug 9, 2007 \XMM\ pointing.\\
$^{c}$ Observed only with \swift. 
Reported value are obtained summing together the last 13 WT \swift\ observations, $N_H=3.2\pm0.2\times10^{22}$ (fixed).\\
$^{d}$ The reported values are from the \CXO\ observation of Jan 25, 2009, the first 
for which there is a partially overlapping in time $INTEGRAL$ observation. 
The PF is calculated over the 0.5$-$10 keV energy range, see \S\ref{sec:timing} for details.\\
$^{e}$ 0.5-10 keV isotropic luminosity, for a distance of 4.5 kpc.
\end{table*}

\begin{figure*}[!]
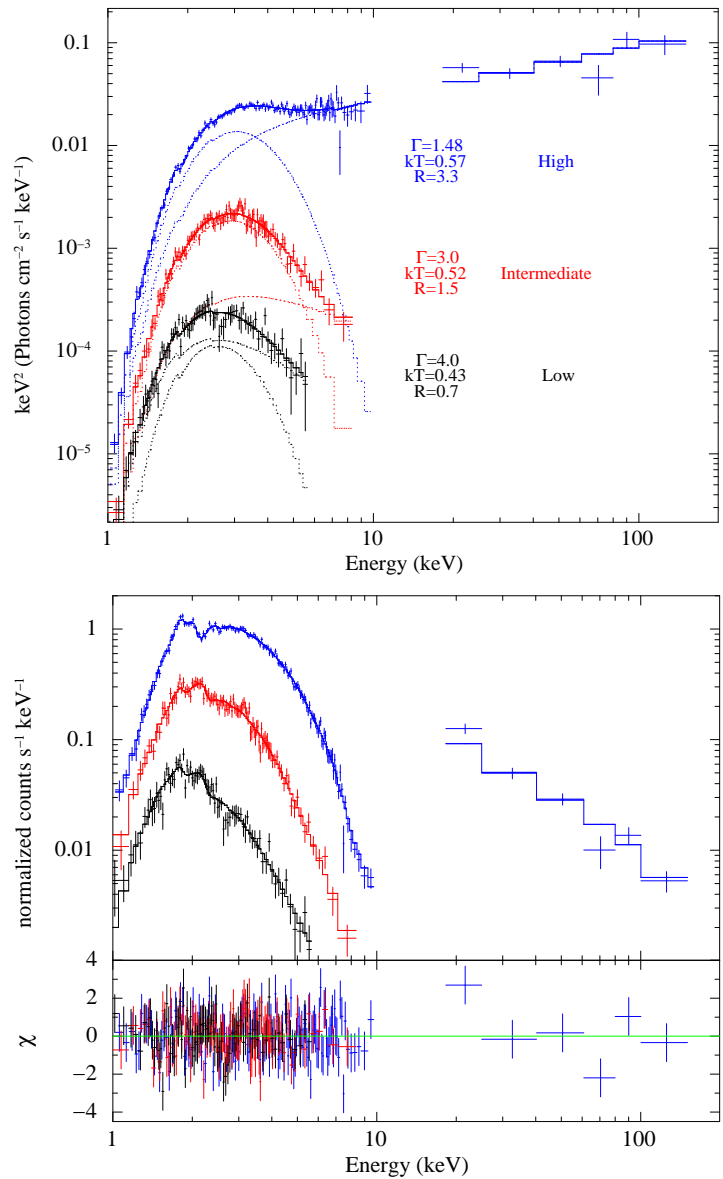

\begin{center}
\begin{tabular}{c}
\includegraphics[angle=-90,scale=.43]{3states.ps}\\
\includegraphics[angle=-90,scale=.45]{norm3states.ps}\\
\end{tabular}
\caption{\textit{Upper panel}: Unfolded source spectra for 
observed intensity levels as modeled with the BB+PL model (best fit parameters are reported in Table 
\ref{tab:activitystate}). High (blue), intermediate (red) and low (black) 
intensity data are from the observation of 
Jan 25, 2009 (\CXO\,), Aug 9, 2007 (\XMM\,), and Aug 21, 2006 (\XMM\,) respectively. 
The high intensity spectrum is the only one for which an $INTEGRAL$ (13--200 keV) 
pointing is available. \textit{Lower panel}: The same as upper panel, but 
for count spectra (residuals are shown in the bottom panel).}
\label{fig:3states}
\end{center}
\end{figure*}

We performed a detailed spectral analysis of the three flux states (high, intermediate and low, 
see section \ref{section:spec} for definition of the three states), which were empirically 
selected from the analysis of figure \ref{fig:fluxhistory}.
In order to compare the spectra of the three recorded flux levels 
%(high, intermediate and low, see fig \ref{fig:fluxhistory}, and section \ref{section:spec} for definition of the three states), 
we used again the BB+PL
model (with absorption) and we carried out a joint fit of the 
Jan 25, 2009 \CXO\ spectrum (high state), Aug 9, 2007 \XMM\ spectrum 
(intermediate state), and Aug 21, 2006 \XMM\ spectrum (low state)
\footnote{The \XMM\ data of August 2006 and August 2007 were
reprocessed using SAS (9.0.0) and the latest calibration files
available.}. Also in this case, we imposed that N$_{H}$ remained the same across all epochs, 
while all other parameters were left free to vary at different epochs ($\chi^2=1.06$ for 313 d.o.f.). 
The results, reported in Table \ref{tab:activitystate}, can be summarized as follows 
(see also Figure \ref{fig:3states}):
\begin{itemize}
 \item \textbf{Low flux level}: The X-ray flux was of
order of $4\times10^{-13}\,\ergscm$ ($L_{4.5\rm\, kpc}\sim9\times10^{32}$ erg s$^{-1}$), 
and the spectrum is described by the sum of a BB of temperature $kT=0.43\pm0.3$ keV and radius
$R=0.7\pm0.2$ km, and a PL with photon index
$\Gamma=4.0\pm0.2$.  Only an upper limit on the PF was
obtained, PF$\lesssim15\%$.\\
 \item \textbf{Intermediate flux level}: The X-ray flux was  
$2-5\times10^{-12}\,\ergscm$ ($L_{4.5\rm\, kpc}\sim5-12\times10^{33}$ erg s$^{-1}$).  The minimum value of
$2\times10^{-12}\,\ergscm$ appears to be well defined by both the
$Einstein$ and $ASCA$ archival data sets (of 1980 and 1998
respectively) and by the latest August-October 2007 \swift\
observations. The spectrum is described by the
sum of a BB with temperature $kT=0.52\pm0.01$ keV and radius
$R=1.5\pm0.1$ km, and a PL with photon index
$\Gamma=3.0\pm0.4$. The PF was $\sim7\%$.\\
 \item \textbf{High flux level}: The X-ray flux varied
between a maximum of $\sim6-8\times10^{-11}\,\ergscm$ 
($L_{4.5\rm\, kpc}\sim1.5-1.9\times10^{35}$ erg s$^{-1}$) recorded during
October 2008 and January 2009, and a minimum which has so far been explored only by \swift\ 
(which did not provide high S/N data). The spectrum was
described by the sum of a BB of average temperature $kT=0.57\pm0.01$
keV and radius $R=3.1\pm0.2$ km, and a PL with photon index
$\Gamma\sim1.5$. The pulsed fraction is highly variable, ranging from
a minimum of $10\%-20\%$ at the highest flux level, to a
maximum of $\sim50\%$ approximatively three weeks later.
The lower level high-flux observation gave $1-1.5\times10^{-11}\,\ergscm$ 
($L_{4.5\rm\, kpc}\sim3\times10^{34}$ erg s$^{-1}$), which
corresponds to the value recorded during both the end of October 2008 \swift\ monitoring
and the September 2009$-$June 2010 \swift\ monitoring.\\
\end{itemize}

The possible final part of the high flux state
($F_{1-10\,\rm keV}=1-1.5\times10^{-11}\,\ergscm$) has
been observed to date only with \swift\ 
(average observation exposure time of about 3 ks). 
%The single pointings do not provide the
%statistics required for a detailed study.  The spectral fit with two
%components (i.e. BB+PL) is not statistically required, making it harder to
%compare these data with those from \CXO\ and \XMM. Consequently, 
In order to improve the S/N, we summed together the last 13 (WT) \swift\ spectra 
\footnote{OBS ID: 00030956046$-$48,51, 53$-$59, 61$-$63.}
(where the X-ray flux and the spectral parameters are 
constant within the uncertainties), 
and performed a fit with the standard BB+PL model. 
Leaving free to vary all the model parameters,
and fixing the column density to an average value 
consistent with the previous analysis, $N_H=3.2\pm0.2\times10^{22}\,\rm cm^{-2}$,
the inclusion of a PL component becomes statistically significant ($P>3\sigma$). 
Its photon index was $\Gamma=5\pm1$, the BB temperature 
$0.69\pm0.02$ keV, and the radius $1.6\pm0.1$ km. 
This finding could suggests that, as the flux decreases during the
high state, the spectrum becomes softer, likely approaching the
intermediate flux state parameter values of both the BB and the PL components. 

We emphasize that the recorded spectral variations, which we supposed to be flux dependent, 
could be time dependent too: we empirically defined three flux states using horizontal lines in figure \ref{fig:fluxhistory}, 
but another possible grouping could be made using vertical lines. We found indications that the source spectrum 
would recover, which is the simplest physical expectation, but current data set can not provide an unambigous confirmation since, 
up to now, we could have observed only one ``cycle'' of variability.

\subsection{Pulse-phase spectroscopy}
\label{sec:pps}

We performed a pulse-phase resolved spectroscopic analysis of
\CXO\ and \XMM\ data. The three \CXO\ pointings without the HETG grating 
were first analyzed separately then summed in order to improve the S/N. 
Both the \XMM\ spectrum and the \CXO\
single and summed-spectra were divided into 4 phase intervals 
(0--0.25, 0.25--0.5, 0.5--0.75, 0.75--1), 
in order to rely upon a large enough number of photons. 
The phase intervals were selected so as to 
separate the two different peaks of the pulse profile
(see Figure \ref{fig:pulse}).
No significant (P$>3\sigma$) changes of the model parameters (BB+PL and NTZ)
were found in both \CXO\ (single and summed spectra) then in \XMM\ data.

\subsubsection{The spectrum in the 0.5$-$200 keV energy range}
\label{subsub:05200}
We then applied the BB+PL spectral model to the whole 0.5--200 keV energy range 
by using data collected by $INTEGRAL$ satellite.
The \INT\ dataset (orbits 767$-$771) was divided into three segments 
in order to carry out spectral fits which overlap (partially) in time 
with \CXO, and \XMM. The first one includes the observations
from Jan 24, 2009 at 16:04 UTC to Jan 25, 2009 at 20:28 UTC, 
for an effective exposure time of 98 ks.
The second time interval starts on Jan 28, 2009
at 15:23 UTC and ends on Feb 01, 2009 at 03:30 UTC,
for an exposure of 191 ks, while the last one 
starts on Feb 1, 2009 at 15:50 UTC,
and ends on Feb 7, 2009 at 05:30 UTC, 
for a total exposure time of 156 ks. 

We checked whether a BB+PL model provides a good fit over the whole
0.5$-$200 keV energy range. The three 0.5$-$10 keV observations of 
Jan 25, Jan 29, and Feb 03, 2009 (with 13--200 
keV $INTEGRAL$ data which partially overlap in time) were fitted individually adopting a BB+PL 
model. All model parameters were left free to vary (with the exception 
of $N_H$ that was kept fixed at $N_H=3.46\times10^{22}\,{\rm cm}^{-2}$ 
see section \ref{sub:bbpl}.).

\begin{figure}
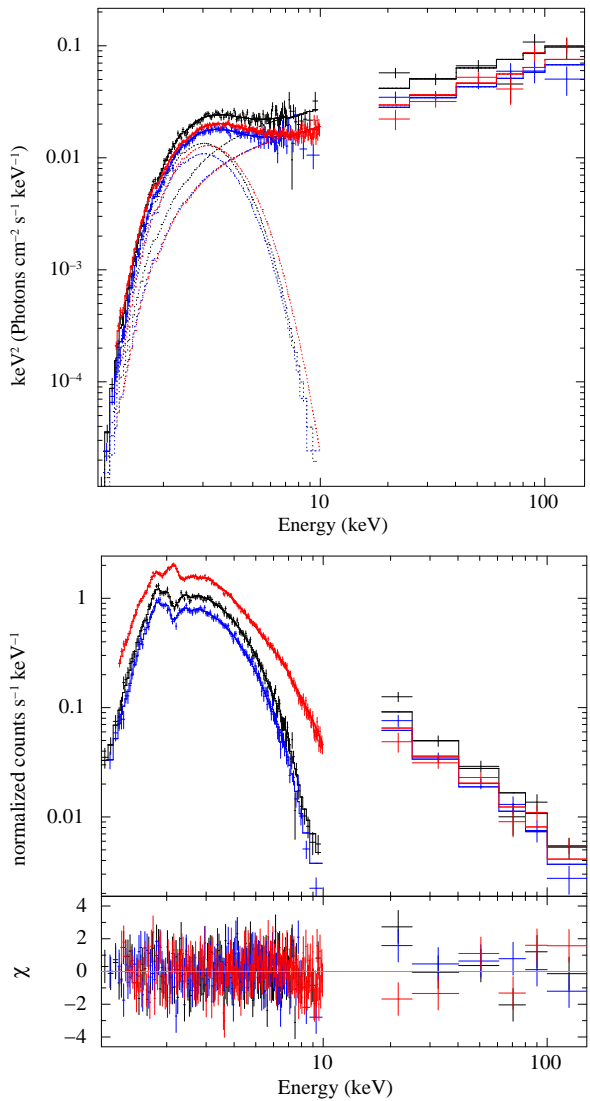

\begin{center}
\begin{tabular}{c}
\includegraphics[angle=-90,scale=.40]{bbplint.ps}\\
\includegraphics[angle=-90,scale=.42]{bbplintnorm.ps}\\
\end{tabular}
\caption{\textit{Upper panel}: Three partially overlapped in time 0.5-10 keV (\CXO\ and \XMM) 
and 13-200 keV ($INTEGRAL$) observations. 
Spectral fits consist of the sum of a BB and a PL  
(black: \CXO\ data of Jan 25, 2009;
blue: \CXO\ data of Jan 29, 2009;
red: \XMM\ data of Feb 03, 2009. 
The same color code applies to the three 
$INTEGRAL$ observations). \textit{Lower panel}: The same as the left panel 
except that count spectra and models are plotted here. Fit residuals are shown in the bottom panel.}
\label{fig:chanint}
\end{center}
\end{figure}

A BB+PL model gave, for the Jan 29, 2009, Feb 03, 2009 and Jan 25, 2009 observations,
a $\chi^2$ value of 0.95 (131 d.o.f.) and 1.00 (205 d.o.f.), and 1.17 (136 d.o.f.) respectively.  
We conclude that the BB+PL model provides a good fit over the whole 0.5$-$200 keV energy range. 
The result of this analysis are reported in Table \ref{tab:tabintegral} 
and Figure \ref{fig:chanint}.
\begin{table*}
\caption{BB+PL spectral parameters in the 0.5--200 keV energy range. The source flux is given separately
in the 0.5--10, and 13--200 keV bands. 
$N_H=3.46\pm0.01\times10^{22}\,{\rm cm}^{-2}$; 1$\sigma$ c.l. uncertainties reported.}
\begin{center}
\begin{tabular}{cccccccc}
\hline \hline
\\
Epoch & $kT_{\rm BB}$  & $R_{\rm BB}$ & $\Gamma $  &  $F_{0.5-10\,keV}$ & $F_{13-200\,keV}$ &$\chi^2$&d.o.f.\\
      &              &               &              &   $\times10^{-11}$ &  $\times10^{-10}$  & &\\
      &   keV       &   km           &             & erg cm$^{-2}$ s$^{-1}$ & erg cm$^{-2}$ s$^{-1}$& &  \\
\hline \hline
\\
Jan 25 2009&  $ 0.56\pm0.01$  &  $  3.46\pm 0.02 $  & $1.50\pm0.03  $  & $5.8\pm0.7$& $3.0\pm0.5$& 1.17& 136 \\
Jan 29 2009&$ 0.56 \pm 0.01  $ &   $  3.08 \pm 0.02   $ & $1.50   \pm 0.03   $  &  $4.2\pm0.4$&$2.1\pm0.1$&  0.95 & 131   \\
Feb 03 2009$^a$ &$ 0.601 \pm 0.004  $ &   $  2.76 \pm 0.02   $ & $1.47    \pm 0.01   $  & $4.5\pm0.1$&$2.3\pm0.3$& 1.00&  205 \\
\\
\hline \hline
\end{tabular}
\label{tab:tabintegral}
\end{center}
$^a$ The 0.5-10 keV spectrum is from \XMM\ data. 
\end{table*}
The average 0.5$-$200 keV spectral index, $\Gamma=1.50\pm0.03$, 
turned out to be slightly harder then that derived from the 0.5$-$10 keV spectra ($\Gamma=1.8\pm0.1$);
however, the two values are consistent to within $3\sigma$.
No softening trend for $\Gamma$ was found in the 0.5-200 keV spectra.
This matches the result of the 0.5--10 keV analysis which showed a variation for $\Gamma$ only 
when comparing the first observation with last one. 

A similar hard PL tail in the 0.5$-$200 keV energy range, was detected
also by Suzaku during a 33 ks observation carried out on January 28$-$29
2010 (Enoto et al. 2010) extending up to 110 keV ($\Gamma^{Suzaku}=1.50\pm^{0.06}_{0.05}$). 
Evidence for the presence of a PL with the same spectral index ($\Gamma\sim1.5$) was
found also by Israel et al. (2010) during the previous outburst of the source
(October 2008). 

During the first days after the 2009 outburst onset, at least up to Feb 03 2009,  
the energy output of \e15\ is dominated by the hard component extending up to 200 keV at least,
indeed the flux in the 13--200 keV range ($3\times10^{-10}\,\ergscm$) is always a factor five higher then in the 0.5--10 keV range 
($6\times10^{-11}\,\ergscm$). 

\subsubsection{Resonant Compton Scattering model}
\label{rcs}
In the following we consider a different modeling of the 0.5--10 keV data 
based on resonant cyclotron scattering, (RCS, Thompson, Lyutikov \&
Kulkarni 2002). In the RCS scenario, the seed photons coming from the
NS surface are up$-$scattered (by multiple consecutive scattering) at
higher energies by electrons and/or positrons populating
the magnetosphere. A semi-analytical treatment of  RCS in 1D was first
developed by Lyutikov and Gavriil (2006), and then successfully 
applied to a large sample of  0.5--10 keV spectra from magnetar candidates by Rea et al. (2008).

We used the NTZ model, a 3D treatment of RCS developed by Nobili,
Zane, Turolla (2008a,b), and already applied to the quiescent emission
of magnetars by Zane et al. (2009), to describe the January outburst of
the transient magnetar \e15. The main model parameters are: 
the value of the twist angle, $\phi$, 
the temperature of the seed blackbody photons T$_{\gamma}$, 
and the bulk motion velocity $\beta_{bulk}$. The polar field strength 
was fixed at $10^{14}$ G, according to the measured $P$ and $\dot{P}$ 
parameters of the source. The NTZ model has the same number of free 
parameters as the standard BB+PL model (this allows for a 
direct comparison of the $\chi^2$ values obtained from the application of the two models). 
In the NTZ model the radius of the emitting region is given by
\begin{equation}
R_{km}=0.78\times(D_{kpc})\times\Big[\frac{N}{(T_{keV})^{3}}\Big]^{\frac{1}{2}}
\end{equation}
where $D_{kpc}$ is the source distance in kpc, $N$ is the model 
normalization, and $T_{keV}$ is the temperature of the seed photons in keV.

As in the case of the BB+PL analysis, the fit was performed
simultaneously on the data of all epochs, by leaving the
parameters free to vary, with the only constraint that the hydrogen
column density be the same at all epochs. Results are reported 
in Figure \ref{fig:speceeufs}, and in Table \ref{tab:spec}. 
The column density derived from the NTZ fits is 
$N_H=3.06\pm0.02\times10^{22}\,{\rm cm}^{-2}$, 
(slightly lower then in the case of the BB+PL analysis). 
The twist angle $\phi$ was found to be constant during the
outburst, to within the uncertainties. 
The average $\phi$ value was $0.48\pm0.01$ rad. 
This result is in agreement both with theoretical expectations 
(Beloborodov 2010) as well as the analysis of the long term evolution
of the transient AXPs \xte\ and \cxou\ (Albano et al. 2010), 
which indicate that the twist angle changes over a timescale of months/years.

A comparison between the low state of activity
\footnote{The NTZ parameters of the August 2006 observation 
are taken from Zane et. al. (2009)} and the outburst revealed that 
as the flux and the radius of the emitting region increased (from
$F^{Aug06}_{1-10\,keV}=3.3\pm^{0.1}_{0.3}\times10^{-13}\,\ergscm$ to
$F^{Jan09}_{1-10\,keV}=6.2\pm0.2\times10^{-11}\,\ergscm$, and from
$R_{Aug06}=2.1\pm0.5$ to $R_{Jan09}=5.2\pm0.5$ km
respectively), $\beta_{bulk}$ and $kT$ also increased, while $\phi$
decreased.  $\beta_{bulk}$ varied from $0.15\pm0.05$ to
$0.72\pm0.02$, $kT$ from $0.38\pm0.01$ keV to $0.69\pm0.02$ keV,
and $\phi$ from $1.14\pm0.08$ rad to $0.48\pm0.01$ rad.
The X-ray flux increase giving rise to the outburst can be, consequently, explained by 
the injection of magnetic energy on the star surface and magnetosphere. 
In fact, we find that both the energy of the charges populating the magnetosphere
and the seed photons temperature increase when the outburst occurs, 
while the twist angle decreases.

As the flux decayed, since Jan 23, 2009, 
all parameters decreased, except for $\phi$ 
(see Table \ref{tab:spec} and Figure \ref{fig:speceeufs}).
The $\chi^{2}_{red}$ of the joint fit was 1.05 for 704 (d.o.f).

The outburst flux decay can be explained
by a decrease in the energy of the charges populating the
magnetosphere, possibly accompanied by a decrease in the size of the
emitting region. We note that the decrease in the twist angle in going 
from the outburst to a less active state appears somehow in contradiction 
with the predictions of the twisted magnetosphere model. In fact the twist 
angle is expected to increase approaching an active state 
(Thompson, Lyutikov \& Kulkarni 2002; see also Mereghetti et al. 2005 for the case of SGR 1806-20). 
A possibility is that the twist was building up while the source was in the 
low/intermediate flux state and then it was in part very quickly dissipated when it entered the outburst state. 

\begin{figure}
\begin{center}
\includegraphics[angle=-90,scale=.45]{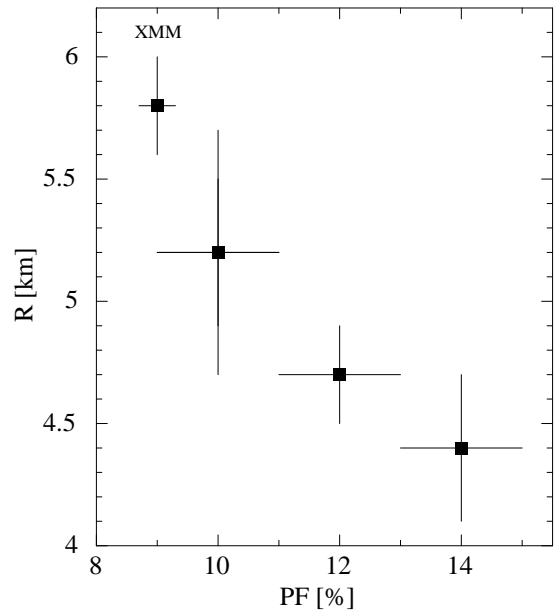}
\caption{Radius of the emitting region, as inferred from the NTZ modeling,
versus the 0.5$-$10~keV PF. See text for details.}
\label{fig:rvspf}
\end{center}
\end{figure}

Similarly to the case of the BB+PL model, 
the application of the NTZ model also suggests that only a part of the NS surface 
is heated and radiates as a hot BB component.
The radius of this region varies (not monotonically) between a maximum 
of 5.8 km (\XMM\ pointing) and a minimum of 4.3 km.
However, the emitting region for the NTZ model 
has a radius of about 5 km compared to $\sim2.5-3$  km  
in the case of the BB+PL model. A possible interpretation 
of the apparently random changes of the
0.5$-$10 keV PF with time is given by the analysis of fig
\ref{fig:rvspf}: higher values of the pulsed fraction are possibly
linked to a shrinking of the emitting region on the star surface. For a given geometric
configuration, the PF increases with the decrease of the emitting
area. 

No fit of the joint \XMM\ and $INTEGRAL$ data could be attempted in this case
because the Montecarlo calculation used to 
tabulate the NTZ model included in $XSPEC$ is based on the non-relativistic resonant
scattering cross section and becomes unreliable above a few tens of keV
(Nobili, Turolla \& Zane 2008a). A more complete treatment, which includes
the full QED cross section, has been presented in Nobili, Turolla \& Zane
(2008b), but no $XSPEC$ model is available for it yet. 

\section{Discussion}
\label{sec:discuss}
The analysis of the whole X-ray data set showed that the source displays 
three different flux levels: low, intermediate and high. By studying the high state, 
which has the high S/N, we were able to find the best spectral model which 
resulted to be the standard (for magnetars) phenomenological 
spectral model composed by a blackbody plus a powerlaw. 
To investigate the variation of the source properties among 
different flux states we also used this phenomenological model. 
However, spectra from the high flux level were also well reproduced 
in terms of a more physical model (NTZ) taking into account the effect of a twisted magnetosphere. 

The comparative analysis of the low, intermediate,
and high flux states of \e15 using the BB+PL model, due to a poor characterization of the low state and to 
a sparse observational coverage, does not provide enough information 
to single out among competitive model, the one that 
should account for the source properties variation 
over the range of observed fluxes (e.g. magnetospheric twist or deep crustal heating). 
However, a trend in the data is clearly present.
The recorded X-ray flux variations from the low to the hight flux state can be simply 
explained by a hardening of the whole 0.5$-$200 keV spectrum (see also Figure \ref{fig:3states} and 
table \ref{tab:activitystate}). According to the BB+PL spectral decomposition, this hardening is due to: 
(a) an increase of the BB temperature from a minimum of
$kT=0.43$ keV to a maximum of $kT=0.57$ keV; 
(b) an increase in the radius of the BB from a minimum of
$0.7$ km to a maximum of $3.3$ km; (c) an hardening of the PL photon index 
from $\Gamma=4.0$ to 1.5. During the high flux level, the hard PL tail with $\Gamma\sim1.5$ is clearly 
extending up to 200 keV (at least), moreover, the flux in the 13--200 keV range is a factor 5 higher then 
that in the 0.5--10 keV range. 

\subsection{Pulsed fraction}
\label{sec:disc}
The analysis of the PF variation with the state of activity is
hampered by the very low S/N ratio 
of the low and intermediate states.
However, by taking as lower limit 
for the PF the value recorded during the intermediate
state ($\sim7\%$, which is fully consistent with the upper limit of
15\% recorded during the low state), it is evident that the PF is
higher during the high state 
(where the PF reached a maximum value of $\sim50\%$).  

The study of the PF vs energy during the 2009 high flux state of \e15\ revealed 
that unlike the majority of the other magnetars, where the periodic 
modulation is higher at higher energies, in the case of \e15 the low energy
band shows the largest level of pulsation. 
At higher energies ($E>4.5$ keV), where the PL dominates and the PF is lower.
These findings suggest that the majority of the modulation comes from the BB component. 
Consequently, the fact that the PF increases, 
from the low to the high flux state, 
is mainly due to the appearance on the NS surface of a hotter  
($kT\sim0.6$ keV) region with radius $\sim3$ km. 

The low level of pulsation recorded for \e15 at 
low and intermediate fluxes ($PF\sim7\%$), given the small radius of the BB region ($\sim1$ km), 
could be explained in terms of a pretty aligned rotator. 
Moreover, when the outburst occur this BB region could increase in size 
up to $R\sim3$ km (as detected during both outbursts), but 
since the geometry is almost aligned the 
pulsed fraction level could remains low ($\sim10-20\%$).

Also the pulse$-$phase spectroscopic analysis corroborates this hypothesis.
The portion of the emitting region on the NS
surface which is in view does not vary significantly as the star rotates, 
resulting in a low level of modulation.
Indeed the radius of the BB responsible for the magnitude of the modulation
is rather high, $R_{BB}\sim3$ km, compared for example to that
measured during the outburst of another transient magnetar, \xte, 
for which $R_{bb}\leq1\,{\rm km}$ (with $kT\sim0.6$ keV) and
the PF was $\gtrsim50\%$ (Bernardini et al. 2009). 
%Alternatively, if the NS radius is small (e.g. $\sim9-10$ km), 
%the gravitational lensing effect is stronger and it 
%could contribute to reduce the PF to the recorded low values (e.g. Nollert et al. 1989).

During the 2008 October outburst, as reported by Israel et al. (2010),
a BB region of about $R\sim3$ km (the same size as the one recorded
during the 2009 January outburst) appeared on the NS surface. In this
case, however, the recorded PF was higher ($\sim20-50\%$),
suggesting that the viewing geometry could be different.
This could suggests that two regions of about the same size, 
were heated after the two different outbursts, 
but their position with respect to the line of sight could be different. 
However, taking into account a longer baseline for the January 2009 
outburst, using the \swift\ data covering the time period between 
January 2009 and June 2010, the PF showed the same evolution in time 
as in the case of the October 2008 outburst. 
Indeed in both cases, after the outburst onset, there was an
anti-correlation between the X-ray flux and the PF;
higher flux levels were associated to lower PFs (see also
Ng et al. 2010). After the October 2008 outburst 
the PF increased from $\sim20\%$ up to 
$\sim50\%$ while after the January 2009 outburst the PF increased 
from $8\pm2\%$ to $33\pm5\%$. 
%In four other magnetar candidates (\e15, \cxou, \sgrn, \sgp; 
%Israel et al. 2007, Rea et al. 2008, Esposito et al. 2008, 
%Esposito et al. 2010a, G{\"o}{\u g}{\"u}{\c s} et al. 2010)
%whenever a long time-scale monitoring of an outburst with adequate S/N was carried out, 
%a clear anti-correlation between PF and flux intensity was seen. 
%The only exception so far is \xte\ (Gotthelf \& Halpern 2007, Bernardini et al. 2009). 
%For \cxou\ this anti-correlation seems to extend all the way to quiescence. 
%Indeed the quiescent PF, $\sim80\%$, dropped to $\sim10\%$ at the outburst 
%onset and increased thereafter during the outburst decay gradually approaching 
%the quiescence PF value (Israel et al. 2007). On the contrary, 
%for \e15 the anti-correlation between PF and X-ray flux seems 
%to hold only during the outburst and does not extend to quiescence 
%(where the PF is lower then during outburst).
However, this anticorrelation seems 
to hold only during the outburst and does not extend to quiescence 
(where the PF is lower, $\lesssim15\%$, then during outburst).

\subsection{Comparing the 2008 and 2009 outburst flux decays}

The best fit model for the flux decay of the 
January 2009 outburst, similar to the October 2008 outburst,
is a PL, $\propto (t-t_{0})^{-\alpha}$, 
but with a higher value of $\alpha$ which resulted to be 0.34 
compared to 0.17 in the case of the previous outburst. 
While the first 2008 \swift\ pointing was carried out only $\sim100$~s 
after the outburst onset, and the recorded flux during this pointing was $\sim6.3\times10^{-11}\,\ergscm$, 
the first \swift\ 2009 pointing was carried out $\sim2$ hours after the 
outburst onset and the recorded flux was $\sim8.0\times10^{-11}\,\ergscm$. 
We conclude that the second outburst 
was more intense then the first, 
and its flux decay was steeper.
%The maximum recorded flux of 
%October 2008 ($F^{Oct08}_{2-10\,keV}=6.3\pm0.5\times10^{-11}\,\ergscm$) 
%can be consequently considered a good approximation of the 
%peak luminosity of the outburst. Indeed the maximum recorded flux of January 2009 
%($F^{Jan09}_{2-10\,keV}=8.0\pm0.3\times10^{-11}\,\ergscm$)
%has to be regarded a lower limit for the outburst peak 
%luminosity.
 
In order to estimate the source flux level before the onset of the January 2009 outburst we 
used the flux decay law found by Israel et al. (2010) for the October 2008 outburst extrapolating 
the X-ray flux value at Jan 22, 2009. 
The extrapolated flux level calculated with this procedure 
resulted to be $\sim1\times10^{-11}\,\ergscm$.

%After the January 2009 outburst the X-ray flux decayed 
%following a PL with $\alpha=0.34$.
%A similar decay has been observed from other transient magnetars, 
%with $\alpha$ ranging from 
%$\sim0.2-0.3$ (in the case of the 2008 outburst from \e15, 
%the 2008 outburst from \sg16, or the outburst from 
%\cxou) to $\alpha\sim0.6-1.2$ (in the case of the 1998 outburst of \sg16\
%and the outburst of \sgp\ respectively). 
%\xte\ and \sgrn\ showed a different behaviour, their X-ray 
%flux decayed exponentially with $\tau_{c}$ of 
%$\sim1$ yr and $\sim24$ days respectively 
%(Gotthelf \& Halpern 2007, Bernardini et al. 2009, and Rea et al. 2009).
%However to properly compare the different outburst evolution 
%of other transient magnetars we first need to further extend 
%the baseline of all observations until the quiescent state is reached. 

Generally a transient magnetar spends the most 
part of time in a steady quiescent flux level, then it 
enters in a active state showing a X-ray flux increase of a factor $\gtrsim100$. 
However, our study showed a peculiar behaviour for \e15\,: 
the X-ray flux can suddenly increase, 
reaching a peak of about $8\times10^{-11}\,\ergscm$ (January 2009), starting from 
a level of about $1\times10^{-11}\,\ergscm$ which is greatly above 
the lower detected level ($\sim4\times10^{-13}\,\ergscm$, August 2006). 
Moreover, the X-ray flux can be above the 
low level, in an intermediate and high state, for a long time (years). 
Finally, \e15\ seems to spend almost consistent amount of time in any logarithmic X-ray flux decade. 
This peculiar behaviour makes difficult to define and identify a ``real'' state of quiescence. 
However, the apparent higher burst active duty cycle for \e15 could be a sampling effect: 
\e15-like-outubrsts from other magnetar candidates could have been missed since the statistics is still fairly poor at present.
In fact, \e15\ is likely one of the transient magnetars 
with the highest number of available observations at different epochs (and flux levels).
  
We remark that an analysis which takes into account the whole available X-ray data set 
should be performed for all transient magnetars in order to unveil their nature. 
%\textbf{Moreover, other magnetar candidates could show a \e15-like behaviour; 
%their outbursts could have been missed i.e. due to 
%sampling effect (statistics is still fairly poor at present).}

\section{Conclusions}
The main results of this work can be summarized as follow:
\begin{itemize}
 \item The analysis of the whole X-ray data archive revealed that the source shows three flux states: 
 low, intermediate, and high. This behaviour, at present unique among transient 
 magnetar candidates, suggests that while not in outburst the source can emit at very different luminosity levels. 
 \item In order to compare the three flux states we used the standard blackbody plus powerlaw model. 
 The spectrum hardens in going from the low to the high state (and vice versa): 
 the powerlaw becomes flatter and the blackbody temperature increases. 
 \item During the high state a powerlaw with spectral index $\sim1.5$ extends without break from 0.5 up to 200 keV (at least) 
 and its flux dominates the source emission. The 13--200 keV flux is a factor 5 higher respect to the 0.5--10 keV flux.
 \item An anti-correlation of the pulsed fraction with the X-ray flux is present during the high flux state 
      (the pulsed fraction is lower when the flux is higher). This anti-correlation does not extend up to the low flux state.
 \item During the high flux state the pulsed fraction decreases with the energy. 
      Most of the periodic modulation is due to the BB component.
 \item We obtained good results also by fitting the high flux state spectrum (that with the higher S/N) 
 with a model (NTZ) which takes into account the effect of resonant Compton scattering in a twisted magnetosphere. 
 This model accounts for the outburst flux increase in term of magnetic energy injection on the star surface 
 and magnetosphere. 
 \item Comparing the two recorded outbursts we found that the peak of the January 2009 outburst is more intense 
 than that of the October 2008 outburst, the X-ray flux decay law is steeper, 
 while the average recorded pulsed fraction is lower. 
 \item We found a unique phase coherent timing solution extending 
 for 15 days after the January 2009 outburst onset. This solution includes P and $\dot{P}$ terms only and 
 consequently resulted to be less complex that the solution found by Israel et al. (2010) 
 extending over 21 days after the onset of the October 2008 outburst.
\end{itemize}

\begin{acknowledgements}

This paper is based on observations obtained with \CXO\, \XMM\, \textit{INTEGRAL}, and \swift.
\CXO\ is part of the Chandra X$-$Ray Center (CXC) which is operated for NASA by the Smithsonian Astrophysical Observatory.
\XMM\ and \textit{INTEGRAL}, are both ESA science missions with instruments and contributions directly funded by ESA Member 
States and the USA (through NASA). \swift\ is a NASA/UK/ASI mission. 
We thank the all duty scientists and science planners for making these
observations possible. F.B acknowledges the \CXO\ help desk. 
PE acknowledges financial support from the Autonomous Region of Sardinia 
through a research grant under the program PO Sardegna FSE 2007--2013, 
L.R. 7/2007 ``Promoting scientific research and innovation technology in Sardinia''.
D.G. acknowledges the French Space Agency (CNES) for financial support.
The Italian authors acknowledge the partial
support from ASI (ASI/INAF contracts I/088/06/0, I/011/07/0, AAE~TH-058,
AAE~DA-044, and AAE~DA-006).

\end{acknowledgements}

\end{document}